\documentclass{emulateapj}
\usepackage{graphicx}
\usepackage{epstopdf}

\shorttitle{RADIO EMISSION FROM $\tau$ BO\"OTIS b}
\shortauthors{Hallinan et al.}

\begin{document}


\title{LOOKING FOR A PULSE: A SEARCH FOR ROTATIONALLY MODULATED RADIO EMISSION FROM THE HOT JUPITER $\tau$ BO\"OTIS b}


\author{G. Hallinan\altaffilmark{1}, S.K. Sirothia\altaffilmark{2}, A. Antonova\altaffilmark{3}, C.H. Ishwara-Chandra\altaffilmark{2},  S. Bourke\altaffilmark{1}, J.G. Doyle\altaffilmark{4}, J. Hartman\altaffilmark{5}  and A. Golden\altaffilmark{6,7}}

\altaffiltext{1}{California Institute of Technology, 1200 E. California Blvd., MC 249-17, Pasadena, CA 91125, USA}
\altaffiltext{2}{National Centre for Radio Astrophysics, TIFR, Post Bag 3, Pune University Campus, Pune 411007, India}
\altaffiltext{3}{Department of Astronomy, St. Kliment Ohridski University of Sofia, 5 James Bourchier Blvd., 1164 Sofia, Bulgaria}
\altaffiltext{4}{Armagh Observatory, College Hill, Armagh BT61 9DG, N. Ireland}
\altaffiltext{5}{Jet Propulsion Laboratory, California Institute of Technology, Pasadena, CA 91109, USA}
\altaffiltext{6}{Department of Genetics, Albert Einstein College of Medicine, Bronx, NY 10461, USA}
\altaffiltext{7}{Centre for Astronomy, National University of Ireland Galway}




\begin{abstract}
Hot Jupiters have been proposed as a likely population of low frequency radio sources due to electron cyclotron maser emission of similar nature to that detected from the auroral regions of magnetized solar system planets.  Such emission will likely be confined to specific ranges of orbital/rotational phase due to a narrowly beamed radiation pattern.  We report on GMRT 150 MHz radio observations of the hot Jupiter $\tau$ Bo\"otis b, consisting of 40 hours carefully scheduled to maximize coverage of the planet's 79.5 hour orbital/rotational period in an effort to detect such rotationally modulated emission. The resulting image is the deepest yet published at these frequencies and leads to a $3 \sigma$ upper limit on the flux density from the planet of $1.2$ mJy, two orders of magnitude lower than predictions derived from scaling laws based on solar system planetary radio emission. This represents the most stringent upper limits for both quiescent and rotationally modulated radio emission from a hot Jupiter yet achieved and suggests that either a)  the magnetic dipole moment of $\tau$ Bo\"otis b is insufficient to generate the surface field strengths of $> 50$ Gauss required for detection at 150 MHz or b) Earth lies outside the beaming pattern of the radio emission from the planet.

\end{abstract}





\section{INTRODUCTION}

The magnetized planets in our solar system produce extremely bright, highly polarized, coherent radio emission at low frequencies, predominantly originating in auroral high magnetic latitudes and attributed to electron cyclotron maser emission (Zarka 1998; Ergun et al. 2000). With the abundant detection of planets around other stellar systems, much theoretical work has been carried out on the possible detection of similar radio emission from extrasolar planets (Zarka et al. 1997, 2001; Farrell et al. 1999; Lazio et al. 2004; Greissmeier et al. 2007; Zarka 2007). As is the case for solar system planets, such a detection represents the most promising method for measuring the magnetic fields possessed by extrasolar planets, in turn providing insight into the composition of the planetary interior. Properties such as planetary rotation rate and inclination of rotational and orbital axes may also be inferred. 

The probability of detection of radio emission from a given extrasolar planet can be crudely estimated from known physical properties of the stellar-planetary system coupled with theoretically derived scaling laws based on solar system planetary radio emissions. One such scaling law, the `radiometric Bode's law' finds the emitted auroral radio power produced by solar system planets to be proportional to the solar wind power incident on the planetary magnetosphere's cross-section (Desch \& Kaiser 1984; Zarka 1992, 1998). Predictions for the expected radio flux from the known populations of extrasolar planets have been derived based on this radiometric Bode's law (Farrell et al. 1999, Lazio 2004), with recent efforts using a more generalized approach considering various forms of interaction between the stellar wind of the host star, the planet and their respective magnetospheres (Zarka 2001, 2007; Stevens 2005, Greissmeier et al. 2007). 

Such studies find that the detection of extrasolar planets with the current generation of ground based radio instruments is strongly biased towards nearby hot Jupiters, with predicted radio power from some planets reaching $10^5$ times that emitted from the Jovian magnetosphere, with resulting flux densities at Earth up to a few hundreds of mJy. In most cases, the primary source of radio power is found to be the incident magnetic energy flux, i.e., the Poynting flux of the interplanetary magnetic field convected on the planet, with lower levels of radio power associated with the kinetic energy flux of the stellar wind, the latter being the dominant source of auroral radio emission from solar system planets. We note that the radiomentric Bode's law assumes that the conversion of kinetic energy flux and and Poynting flux to emitted radio power occurs with the same efficiency for hot Jupiters as for solar system planets, i.e., it does not take into account how this energy conversion efficiency might vary due to the vastly different local plasma conditions for hot Jupiters. 

More recently, Nichols (2011) have argued that rapidly rotating giant planets with plasma sources embedded within their magnetosphere (eg. volcanic moons) and subject to high X-ray/UV illumination from their parent star can produce comparably luminous radio flux at much larger orbits (many AU), associated with magnetosphere-ionosphere coupling. 

Initial attempts to detect auroral radio emission pre-date the discovery of a known population of extrasolar planets and thus necessarily involved blind searches of nearby stellar systems. More recent efforts have predominantly involved observations of nearby hot Jupiters previously detected through radial velocity and transiting observations. As of yet, no confirmed detection has been reported in the literature (Zarka et al. 1997, Bastian et al. 2000, Lazio et al. 2004, Majid et al. 2005). 

\section{The Implications of Narrowly Beamed Emission}
To date, searches for radio emission from hot Jupiters have typically consisted of short duration observations of a small sample of exemplary hot Jupiters, a point that becomes particularly significant when evaluated in the context of the geometrical selection effect associated with planetary radio emission. A signature property of the coherent electron cyclotron emission produced by magnetized planets is narrow beaming at large angles to the local magnetic field (Zarka 1998). The effect is two-fold, a) planets do not beam to 4$\pi$ sr and thus only a fraction will be detectable as radio sources and b) radio emitting planets will only be detectable over ranges of rotational/orbital phase during which the magnetic field in the source region is suitably orientated relative to our line of sight.  The solid angle of the beam of the emitted radiation has been inferred for many of the components of Jovian radio emission, for example, with $\theta = 1.6$ sr for the non-Io decametric component (Zarka et al. 2004).  This narrow beaming accounts for the strong dependence on rotational phase observed for Jovian decametric emission and is also well demonstrated in the unusual properties of the radio light curves observed for a number of very low mass stars and brown dwarfs (collectively referred to as ultracool dwarfs). The latter are found to produce $100\%$ circularly polarized radio emission with narrow duty cycle ($< 10\%$) and pulsed on the rotation period with p $\lesssim 3$ hours for all detected sources thus far (Hallinan et al. 2007, 2008, Berger et al. 2009). These periodic pulses are attributed to electron cyclotron maser emission of fundamentally similar nature to planetary auroral radio emission, albeit detected at GHz, rather than MHz frequencies due to the much stronger magnetic field strengths possessed by ultracool dwarfs. 

It can be expected, therefore, that the radio emission from extrasolar planets may be similarly pulsed, a point further emphasized by recent simulations of the expected dynamic spectra for hot Jupiters, which indicate that detectable emission may indeed be confined to a few percent of rotational or orbital phase (Hess \& Zarka 2011). In the event of such pulsed emission, a robust investigatory observation would require good sampling of the rotational and/or orbital period. However, most hot Jupiters predicted to have detectable radio emission have semi-major axes $< 0.1$ AU  (Greissmeier et al. 2007), such that tidal synchronization of rotational and orbital period can be assumed, with the resulting rotational/orbital periods ranging from $1 - 5$ days. Therefore, previous observations of hot Jupiters, typically of order a few hours in duration, poorly sample the rotational phase of such planets. We have commenced a targeted radio campaign of a small sample of hot Jupiters, specifically scheduled to ensure maximal rotational/orbital phase coverage, in an attempt to detect the putative pulsed emission associated with narrowly beamed electron cyclotron maser emission. The first target in our search is $\tau$ Bo\"otis b.

\section{$\tau$ Bo\"otis b}
$\tau$ Boo b is a hot Jupiter ($M \sin i = 4.1 \pm 0.153 M_{J}$; semi-major axis = 0.0480\,AU; orbital period = 79.5 hours) orbiting a main sequence F7V star at a distance of 15.6 pc (Butler et al. 1997). Multi-epoch spectropolarimetric observations of the host star confirm the presence of a large-scale magnetic field with maximum intensity 5-10 Gauss that undergoes periodic polarity reversal every $\sim 2$ years, similar to that associated with the Solar magnetic cycle (Catala et al. 2007; Donati et al. 2008; Fares et al 2009). These data also suggest that the planet has tidally synchronized intermediate latitudes of the shallow outer convective envelope of the star to the planetary orbital period.

Recent high resolution spectroscopy of carbon monoxide absorption lines in the atmosphere of $\tau$ Boo b have yielded direct measurements of the radial velocity of the planet, thereby breaking the degeneracy in previous $M \sin i$ measurements and allowing determination of both the mass of the planet ($5.6 \pm 0.7 M_{Jup}$) and the inclination of the planetary system relative to our line of sight $i = 47^{+7}_{-6}$ degrees (Rodler et al. 2012). These results were consistent with previous estimates derived from the spectropolarimetric data  of Donati et al. (2008). 

\par With its close proximity, large mass and small semi-major axis, $\tau$ Boo b has been identified as one of the most promising candidates for low frequency radio emission and has been the subject of more observations than any other planetary system. Predicted flux densities inferred through application of the radiometric Bode's law are of order a few hundred mJy with emission powered by incident magnetic energy flux of the interplanetary magnetic field on the planetary magnetosphere (Stevens 2005, Greissmeier et al. 2007). 
\section{OBSERVATIONS}

We used the Giant Metrewave Radio Telescope (GMRT: Swarup et al 1991) at 153 MHz to conduct 5 individual observations of $\tau$ Boo b of 8 hours duration each and scheduled on successive nights from Feb 14-18 2008. We assume tidal locking of $\tau$ Bo\"otis b to its parent star and therefore scheduled observations to ensure there was no overlap in orbital phase coverage of the planet. This led to $\sim 50\%$ total coverage of the 79.5 hour putative rotational period. 

The bandwidth of our observations was 8 MHz and the extragalactic sources 3C286 and 1330+251 were used for flux and phase calibration respectively. Data reduction was carried out using AIPS++. Bad data were flagged throughout the calibration process, including data for antennas with high errors in antenna-based solutions and data affected by radio frequency interference, as identified by a median filter with a $6 \sigma$ threshold. Corrections for variations of system temperature and primary beam between antennas were also applied.  

The image produced through reduction of these data is the deepest integration at 150 MHz yet published in the literature,  with RMS noise of 350 $\mu$Jy throughout most of the image (Figure 1). No source is detected at the position of the planet with a $3 \sigma$ upper limit on the flux density of $1.2$ mJy. We note a slightly increased RMS noise at the location of planet of $400 \mu$Jy due to the sidelobes of two strong sources in the field. We also imaged each individual 8 hour observing block to account for the possibility of emission confined to a narrow range of rotational phase, with no evidence for any detection in the resulting images with RMS noise of $\sim 1$ mJy. 

To further investigate the possibility of narrow duty cycle emission, we created time series at the position of $\tau$ Bo\"otis b. Light curves for a synthesized beam size region at $\alpha_0, \delta_0$ were generated after the final imaging stage. Model visibilities for the entire field of view (excluding the beam wide region at $\alpha_0, \delta_0$) were subtracted out from the final calibrated data. These residual visibilities data (RVD) were then phase centered to $\alpha_0, \delta_0$. All RVD were averaged for desired time bins to obtain light curves. No evidence of heightened emission is present in the time series at any point in the observed rotational phase of the planet. As the radio emission from the planet is expected to be highly circularly or elliptically polarized, time series for the left and right circularly polarized emission are presented in Figure 2 and 3 respectively. 

 \begin{figure}
\epsscale{1.33}
\plotone{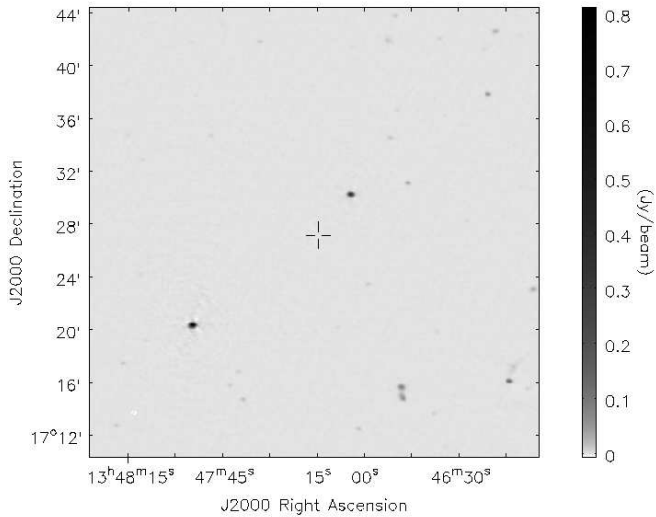}
\caption{The 153 MHz image produced for the integrated 40 hours of observations of $\tau$ Bo\"otis b. The crosshairs indicate the position of the planet. The RMS noise in most of the image is $350 \mu$Jy but rises to $400 \mu$Jy near the position of the planet due to the sidelobes of two nearby sources. No source is detected at the position of the planet.}

\end{figure}

 \begin{figure}
 \epsscale{0.9}
\includegraphics[scale=0.35]{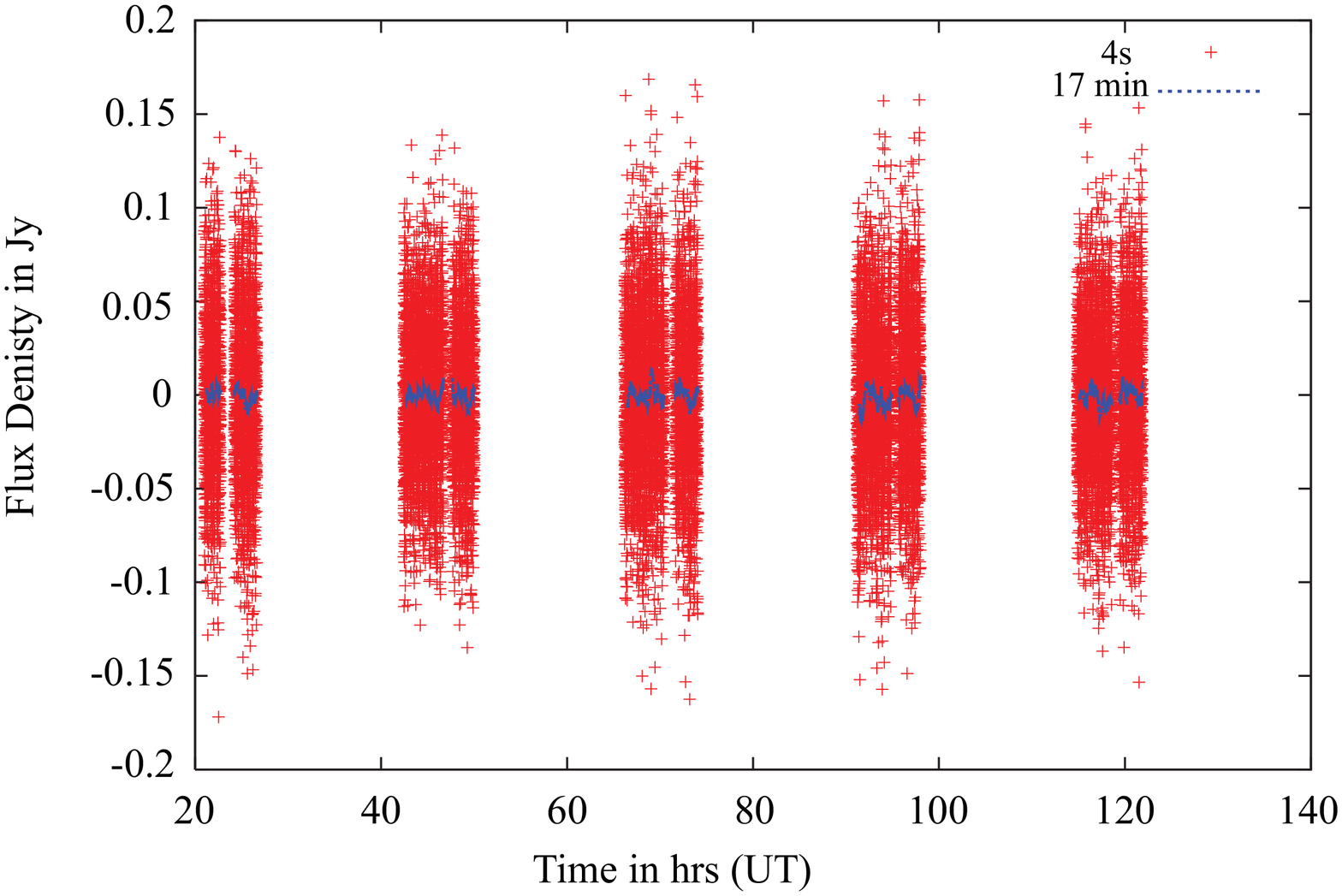}
\caption{Light curves for the left circularly polarized flux at the position of $\tau$ Bo\"otis b over the course of the observing campaign. Time series at time resolution of 4 seconds and 17 minutes are shown in red and blue respectively, with RMS noise of 65 mJy for the 4 second resolution time series and 4 mJy for the 17 minute resolution time series. No evidence for heightened emission is present during any of the observed orbital/rotational phase of the planet.}

\end{figure}

 \begin{figure}
\epsscale{0.9}
\includegraphics[scale=0.35]{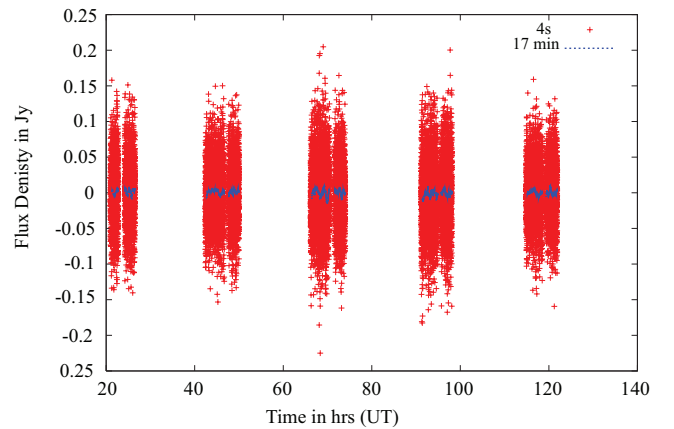}
\caption{Light curves for the right circularly polarized flux at the position of $\tau$ Bo\"otis b over the course of the observing campaign. Time series at time resolution of 4 seconds and 17 minutes are shown in red and blue respectively, with RMS noise of 65 mJy for the 4 second resolution time series and 4 mJy for the 17 minute resolution time series. No evidence for heightened emission is present during any of the observed orbital/rotational phase of the planet.}

\end{figure}

\section{DISCUSSION}

\subsection{The Magnetic Field of $\tau$ Boo b}
A critical factor governing the detectability of the radio emission from $\tau$ Boo b is the strength and extent of the planetary magnetosphere, the former governing the range of frequencies over which the emission can be detected. Electron cyclotron maser emission is produced at the electron cyclotron frequency, where $f_{ce} = eB/2 \pi m_e \approx 2.8 \times 10^6 B$ Hz, and thus planetary radio emission should exhibit a high frequency cut-off corresponding to the electron cyclotron frequency in regions of highest magnetic field strength near the planetary surface, where local plasma conditions are such that the emission mechanism can operate (plasma frequency $f_{pe} \ll f_{ce}$). For example, Jovian decametric radio emission exhibits a cut-off near 40 MHz, corresponding to a maximum magnetic field strength of $\sim 14$ Gauss near the polar regions. 

\par Similarly, one can expect a cut-off frequency for the radio emission produced by extrasolar planets, determined by the maximum magnetic field strength near the surface of the planet. There have been efforts to predict exoplanetary magnetic dipole moments, through extrapolation of scaling laws derived from solar system planets and based on estimates of the relevant properties of the planet, such as the size, density and rotation rate of the convective dynamo region. A number of such models assume a force balance between Coriolis and Lorentz forces, with a resulting strong dependence on rotation rate (Greissmeier et al. 2005, and references therein). This is particularly impactful on the predicted magnetic dipole moments for hot Jupiters, where tidal locking can result in rotation periods of a few days and correspondingly weak surface field strengths. 

\par Taking the specific example of $\tau$ Boo b, surface magnetic field strengths of a few Gauss have been predicted, corresponding to a maximum cut-off frequency well below that targeted in our observations (Greissmeier et al. 2007). However, Reiners \& Christensen (2010) have proposed that the energy flux available at the planetary core determines the planetary magnetic field strength, with much weaker dependence on rotation rate. This model predicts much higher magnetic field strengths for $\tau$ Boo b, such that detection in our observing band of 150 MHz is possible. 

While the stringent upper limit on radio flux from $\tau$ Boo b presented by our observations may suggest that the cut-off in emission for the planet is $< 150$ MHz, data on a much larger sample of planets is required to firmly constrain dynamo theory. Indeed, the discrepancies in the predictive power of current models largely motivates ongoing efforts to detect exoplanetary radio emission. All models do agree, however, that the population of detectable planets should increase with lower frequency down to $\sim 10$ MHz, below which the Earth's ionosphere inhibits ground-based observations. This is consistent with observations of solar system planets, where various components of emission span a large bandwidth ($\Delta f \sim f$) below the cut-off frequency.

\subsection{The Beaming Pattern of the Radio Emission from $\tau$ Boo b}
Observations of the two classes of objects known to emit auroral radio emission, solar system planets and ultracool dwarfs, suggest that emission is confined to specific ranges of rotational phase, consistent with the strong anisotropic beaming of electron cyclotron maser emission. With $50\%$ of rotational phase coverage achieved, it may be possible that our observations simply did not cover the range of rotational phase during which similar pulsed emission from $\tau$ Boo b may be detectable, which would limit the duty cycle of any pulses to $< 20\%$. Pulse profiles with duty cycle $<5\%$ have been observed for the pulsed radio emission from ultracool dwarfs, making this a plausible scenario (Hallinan et al. 2007). 

Alternatively, the beamed radio emission might never sweep in in the direction of Earth during a full rotation of the planet, due to the combination of anisotropic beaming and an unfavorable inclination angle of the planetary rotation axis relative to our line of sight. This is observed for solar system planetary radio sources; Cassini observations of Saturn kilometric radiation revealed a beaming pattern that rarely illuminates regions $> 45 \arcdeg$ from the planetary equatorial plane (Lamy et al. 2008). A similar geometrical effect has been proposed to account for the small fraction of ultracool dwarfs detected at radio frequencies (Hallinan et al. 2008). 

\par The beaming pattern of the radio emission from extrasolar planets has been specifically investigated by Hess \& Zarka (2011), who find that virtually no emission is detectable for an inclination of the planetary orbital plane $\geq 60 \arcdeg$, independent of the inclination angle of the associated magnetic field. The recent confirmation by Rodler et al. (2012) of an inclination angle $i = 47^{+7}_{-6} \arcdeg$ for $\tau$ Boo b suggests that an unfavourable beaming pattern is a likely reason for the non-detection of auroral radio emission from the planet.

\section{CONCLUSION}
We have conducted monitoring observations of the hot Jupiter $\tau$ Boo b which yield $3 \sigma$ upper limits on the flux density from the planet of $1.2$ mJy, two orders of magnitude lower than predictions based on scaling laws from solar system planets (Stevens 2005; Greissmeier et al. 2007). Time series analysis does not reveal any transient emission consistent with narrowly beamed electron cyclotron maser emission, making these observations the most stringent upper limits for both quiescent and rotationally modulated radio emission from a hot Jupiter yet achieved. 

\par Future searches for radio emission from extrasolar planets will require extensive monitoring of a large enough population of planets to account for the geometrical selection effect implied by beaming of the emission to $<< 4$ sr, with the necessary sensitivity for detection (a few mJy). Observations should focus on both hot Jupiters as well as planets at larger orbits, the latter of which have been shown to be able to produce detectable levels of emission (Nichols 2011) and to possess higher strength magnetic fields due to higher rotational velocities allowed by the absence of tidal locking. In all cases, full and repeated rotational phase coverage should be maintained to account for the narrow beaming of electron cyclotron maser emission with greater chance of success expected at lower frequencies. The recent detection of rado emission from a very cool 900K T6.5 dwarf (Route \& Wolszczan 2012) certainly bodes well for these efforts and already provides a significant benchmark to constrain dynamo theory in the mass and temperature range between low mass stars and planets.

\par A new generation of low frequency radio telescopes have commenced science observing, or will do so in the near future, including the Low Frequency Array (LOFAR) and the Long Wavelength Array (LWA), while existing arrays, such as the GMRT, JVLA and UTR-2 are undergoing substantial upgrades to their existing low frequency systems. Such instruments will allow a definitive investigation into the radio emission from hot Jupiters possibly opening up a new field of research, exoplanetary magnetospheric physics.

\section{ACKNOWLEDGMENTS}

We thank the staff of the GMRT for their assistance with this observing program. GMRT is run by the National Centre for Radio Astrophysics of the Tata Institute of Fundamental Research. Armagh Observatory is grant aided by the N. Ireland Dept. of Culture, Arts \& Leisure. AA gratefully acknowledges the support of the Bulgarian
National Science Fund (contract No DDVU02/40/2010). JGD wishes to thank the Leverhulme Trust for funding. AG acknowledges support from Science Foundation Ireland (Grant Number 07/RFP/PHYF553).

\end{document}